\documentclass[aps,pra,twocolumn,showpacs,groupedaddress]{revtex4-1}
\usepackage{amsmath}

\usepackage{amsthm}
\usepackage[english]{babel}
\usepackage[T1]{fontenc}
\usepackage[latin1]{inputenc}
\usepackage{graphicx}
\usepackage{amssymb}
\usepackage{color}

\def\r {{\bf r}}

\begin{document}

\title{Reply to ``Comment on: `Single-shot simulations of dynamic quantum many-body systems' [arXiv:1610.07633]''}

\author{Kaspar Sakmann$^{1,2}$\footnote{E-mail: kaspar.sakmann@gmail.com}, 
 and Mark Kasevich$^{1}$}

\affiliation{$^1$ Department of Physics, Stanford University, Stanford, California 94305, USA \\
$^2$ Vienna Center for Quantum Science and Technology, Atominstitut, TU Wien, Stadionallee 2, 1020 Vienna, Austria}

\maketitle

In \cite{SakmannKasevich} we provided an algorithm that simulates single experimental shots based exclusively 
on the postulates of quantum mechanics, i.e. 
the algorithm draws random deviates from the $N$-particle 
probability distribution $P(\r_1,\dots,\r_N)=\vert\Psi\vert^2$, where $\Psi$ is an $N$-boson wave function. 
The algorithm is independent of the method to obtain the wave function. 
We used wave functions obtained from numerical solutions 
of the many-body Schr\"odinger equation using the MCTDHB method \cite{MCTDHB2}.
Very recently, our algorithm was used to identify fragmented superradiance 
of a Bose-Einstein condensate in a cavity \cite{FRAG_SR}.

Phase space methods and approximations are not the topic of our work \cite{SakmannKasevich}. 
We merely mentioned in the introductory paragraph that some researchers 
had attempted to interpret trajectories of the semiclassical truncated Wigner approximation as single shots, 
provided the Wigner function of the state is strictly positive.
This ``single-shot interpretation'' of truncated Wigner trajectories is motivated by interpreting the Wigner function 
as a phase space probability distribution
in direct analogy to classical statistical mechanics. It is thus 
not based on quantum mechanics and well-known inconsistencies arise from it.
While not related to our work, we briefly review some relevant known facts.

One, the ``single shot interpretation''  is a postulate independent of the approximations 
made in the derivation of the truncated Wigner approximation, 
in contrast to what Drummond and Brand claim \cite{DrummondBrand}, 
please see \cite{Blakie} for details. 

Two, states of $N$-bosons have non-positive Wigner functions \cite{Hudson} 
which makes an interpretation as a probability distribution problematic, if not impossible \cite{Blakie, Polkovnikov}.  
In fact in the truncated Wigner approximation not even an $N$-boson wave function appears.
Some researchers adopt the ``single shot interpretation'' but require the Wigner function to be strictly positive \cite{Blakie}.
Others state clearly that ``the individual stochastic realizations or phase-space trajectories of
the complex field amplitude do not have any correspondence to physical observables, 
except in the mean where they correspond to expectation values of symmetrically ordered creation and annihilation operators'' \cite{LewisSwan}.  

Three, even if the ``single shot interpretation''  is  taken as a purely heuristic procedure of simulating single-shots
of practically fully condensed, i.e. single-mode states with positive Wigner functions, 
the results can be inaccurate \cite{LewisSwan}.

We remind the reader again that phase space methods are not the topic of our work. 
Our sole contribution in \cite{SakmannKasevich}
to the above discussion is the explicit analytical demonstration in Supplementary Information section IV
that single shots cannot be simulated from mixed state density matrices, thus ruling out also positive Wigner functions.
Summarizing, the context that Drummond and Brand provide is inaccurate and misleading. 
A ``single shot interpretation'' of truncated Wigner trajectories is not reconcilable with quantum mechanics. 
Although obvious, 
the additional criticism Drummond and Brand put forward on our work (which we refute in what follows) 
cannot cure the above documented problems with the truncated Wigner approximantion method.

Moreover, many interesting phenomena occur when the bosons are not in a single-mode condensed, 
but rather in correlated, fragmented states.
These are beyond the scope of semiclassical approximations such as the truncated Wigner approximation, see \cite{Polkovnikov}.
We consider specifically such many-body systems in our work, 
building on the pioneering works of Javanainen and Yoo \cite{JavYoo} as well as Castin and Dalibard \cite{CasDal} for non-interacting particles.
Doing so, for instance removes the need to introduce a hidden variable for the explanation of the 
appearance of interference fringes in the collision of independent condensates \cite{MulLal}.
This should already clarify the scientific value of single-shot simulations of pure states that Drummond and Brand question in their correspondence.

Furthermore, Drummond and Brand's criticism of the MCTDHB method is unfounded.
By systematically including more and more orbitals, 
the MCTDHB method converges to the exact solution of the many-body Schr\"odinger equation.
In \cite{HIM} MCTDHB was benchmarked against an exact analytical solution of the interacting $N$-boson Schr\"odinger equation.  
Ground-state energies were obtained with an impressive accuracy of $11$(!) digits for $N=1000$ and $M=3$ orbitals, 
as well as $15$(!) digits for $N=10$ bosons and $M=7$ orbitals \cite{HIM}. 
Nonequilibrium dynamics was also benchmarked in the same work. 
We are not aware of any other method that provides results of comparable accuracy on this benchmark.
Additionally, in Fig.~3 of our work \cite{SakmannKasevich} we demonstrate that numerically-exact results
can even be obtained for the full distribution function of $N$-body operators using MCTDHB. 
The exact analytical result and the numerical one are indistiguishable.

For multiconfigurational time-dependent Hartree methods \cite{MCTDH,Manthe92,MCTDH_book} 
convergence was proven in general \cite{Lubich}, 
invalidating Drummond and Brand's claim that convergence is merely ``conjectured''.
Thus, the picture that Drummond and Brand try to convey of the MCTDHB method is incorrect and misleading. 
MCTDHB is a very successful method based on the variational principle \cite{Dirac,Frenkel}
for obtaining highly accurate, sometimes even numerically exact solutions of  the time-dependent $N$-boson Schr\"odinger equation, 
see e.g. \cite{BJJexact,HIM}. 

In the specific examples shown in Fig.~1 and Fig.~2 it was neither necessary nor intended 
to fully converge to numerically exact results, as done in Fig.~3. 
In Fig.~1 we analyze the collision between two independent attractively interacting BECs. 
In the simplest approximation two independent BECs are described by $M=2$ orbitals. 
We found that during the collision additional natural orbitals get populated and make it 
impossible to separate the condensates after the collision into two individual subsystems, 
as would be the case for noninteracting systems. 
$M=3$ is the minimal number of orbitals that is needed to establish this result. 
We used $M=4$.  

In Fig.~2 we considered vortex formation which is usually described on the basis of Gross-Pitaevskii mean-field 
theory, i.e. MCTDHB using $M=1$ orbitals. 
We wondered whether mean-field theory would be enough to describe the formation of vortices. 
To answer this question at least $M=2$ orbitals are needed. 
Using $M=2$ orbitals we found that here 
vortex formation is a many-body phenomenon and not a mean-field process. 
Single shots revealed that from some point onwards one would find vortices at fluctuating positions.
Very recently, these findings were supported in \cite{Phantom} with the use of more orbitals.
 
We did not claim anywhere that these two examples were numerically exact solutions of the $N$-boson Schr\"odinger equation, 
but this was also not necessary. The variational principle ensures that using more orbitals cannot change the above conclusions. 
Of course it would be possible to obtain numerically exact results for fewer particles, but in these two examples we 
preferred to use parameter values that are closer to current experiments instead of restricting ourselves exclusively to numerically exact results.

Finally, Drummond and Brand raise interesting questions regarding the status of current theory and its role in attempting to understand the phenomenology of modern many-body systems which are now becoming experimentally accessible, and which are computationally difficult. 
Progress in understanding these systems will likely require the interplay between theory and experiment. 
The phenomenology of some of these systems might only be captured using quantum algorithms on quantum simulators. 
An interesting question is the role that beyond-mean-field algorithms, which capture some many-body correlations, can play. 
It is our hope that the single shot algorithm introduced in our work will do so.
The recent investigations in \cite{FRAG_SR} make us optimistic that this will indeed be the case.  

Drummond and Brand also question the utility of pure state methods, since realistic experiments -- they argue -- necessarily involve 
coupling to an environment and hence demand a density matrix approach.
This perspective is undermined by modern ultracold atom experiments, where the objective is to find regimes where 
such environmental couplings are negligible. 
On the contrary, an exciting frontier are experimental systems where pure-state many-body physics drives the phenomenology
instead.

\end{document}